\documentclass[final,journal,letterpaper,twoside,twocolumn]{IEEEtran}

\ifCLASSINFOpdf
  \usepackage[pdftex]{graphicx}
  \graphicspath{{./figures/}}
  \DeclareGraphicsExtensions{.eps,.pdf,.jpeg,.png,.tif}
\else

\fi

\usepackage{multirow}

\usepackage{amsmath, amssymb}

\usepackage{algorithm,algorithmic}

\usepackage{booktabs}

\usepackage[table,xcdraw]{xcolor}

\usepackage{xcolor}

\usepackage{array}

\ifCLASSOPTIONcompsoc
  \usepackage[caption=false,font=normalsize,labelfont=sf,textfont=sf]{subfig}
\else
  \usepackage[caption=false,font=footnotesize]{subfig}
\fi

\begin{document}
%
\title{SAR Tomography at the Limit: Building Height Reconstruction Using Only 3 -- 5 TanDEM-X Bistatic Interferograms}

\author{Yilei~Shi,~\IEEEmembership{Member,~IEEE},
        Richard~Bamler,~\IEEEmembership{Fellow,~IEEE},
        Yuanyuan~Wang,~\IEEEmembership{Member,~IEEE}, \newline
        and~Xiao Xiang~Zhu,~\IEEEmembership{Senior~Member,~IEEE}
\thanks{This work is supported by the European Research Council (ERC) under the European Union's Horizon 2020 research and innovation programme (grant agreement no. ERC-2016-StG-714087, acronym: So2Sat, www.so2sat.eu), the Helmholtz Association under the framework of the Young Investigators Group ``SiPEO" (VH-NG-1018, www.sipeo.bgu.tum.de), Munich Aerospace e.V. – Fakult{\"a}t f{\"u}r Luft- und Raumfahrt, and the Bavaria California Technology Center (Project: Large-Scale Problems in Earth Observation). The authors thank the Gauss Centre for Supercomputing (GCS) e.V. for funding this project by providing computing time on the GCS Supercomputer SuperMUC at the Leibniz Supercomputing Centre (LRZ).}

\thanks{Y.~Shi is with the Chair of Remote Sensing Technology, Technische Universit{\"a}t M{\"u}nchen (TUM), 80333 Munich, Germany (e-mail: yilei.shi@tum.de)}
\thanks{R.~Bamler is with the Remote Sensing Technology Institute (IMF), German Aerospace Center (DLR), 82234 We\ss ling, Germany and the Chair of Remote Sensing Technology, Technische Universit{\"a}t M{\"u}nchen (TUM), 80333 Munich, Germany (e-mail: richard.bamler@dlr.de)}
\thanks{Y.~Wang is with Signal Processing in Earth Observation, Technische Universit{\"a}t M{\"u}nchen (TUM), 80333 Munich, Germany (e-mail: yuanyuan.wang@dlr.de)}
\thanks{X.X.~Zhu is with the Remote Sensing Technology Institute (IMF), German Aerospace Center (DLR), 82234 We\ss ling, Germany
 and Signal Processing in Earth Observation, Technische Universit{\"a}t M{\"u}nchen (TUM), 80333 Munich, Germany (e-mail: xiaoxiang.zhu@dlr.de)}
 \thanks{\emph{(Correspondence: Xiao Xiang Zhu)}}
}

\markboth{submitted to IEEE TRANSACTIONS ON GEOSCIENCE AND REMOTE SENSING, 2019}%
{Y. Shi \MakeLowercase{\textit{et al.}}: SAR Tomography at the Limit: Building Height Reconstruction Using Only 3 -- 5 TanDEM-X Bistatic Interferograms}

\maketitle

\begin{abstract}
\textcolor{blue}{This is the preprint version, to read the final version please go to IEEE Transactions on Geoscience and Remote Sensing on IEEE Xplore.} Multi-baseline interferometric synthetic aperture radar (InSAR) techniques are effective approaches for retrieving the 3-D information of urban areas. In order to obtain a plausible reconstruction, it is necessary to use more than twenty interferograms. Hence, these methods are commonly not appropriate for large-scale 3-D urban mapping using TanDEM-X data where only a few acquisitions are available in average for each city. This work proposes a new SAR tomographic processing framework to work with those extremely small stacks, which integrates the non-local filtering into SAR tomography inversion. The applicability of the algorithm is demonstrated using a TanDEM-X multi-baseline stack with 5 bistatic interferograms over the whole city of Munich, Germany. Systematic comparison of our result with TanDEM-X raw digital elevation models (DEM) and airborne LiDAR data shows that the relative height accuracy of two third buildings is within two meters, which outperforms the TanDEM-X raw DEM. The promising performance of the proposed algorithm paved the first step towards high quality large-scale 3-D urban mapping.
\end{abstract}

\begin{IEEEkeywords}
SAR interferometric (InSAR), SAR tomography (TomoSAR), TanDEM-X, digital elevation models (DEM), 3-D urban mapping
\end{IEEEkeywords}

\section{Introduction}

\subsection{The TanDEM-X Mission}
\IEEEPARstart{T}{anDEM-X} satellite is a German civil and commercial high-resolution synthetic aperture radar (SAR) satellite which has almost identical configuration as its 'sister' TerraSAR-X satellite. Together with TerraSAR-X, they are aiming to provide a global high-resolution digital elevation model (DEM) \cite{bib:krieger2007tandem}. Both satellites use a spiral orbit constellation to fly in tight formation in order to acquire the image pair simultaneously, which significantly reduces the temporal decorrelation error and the atmospheric interference. Since its launch in 2010, TanDEM-X has been continuously providing high quality bistatic interferograms that are nearly free from deformation, atmosphere and temporal decorrelation.

\subsection{SAR Tomography Techniques}
Tomographic synthetic aperture radar (TomoSAR) is a cutting-edge SAR interferemetric technique that is capable of reconstructing the 3-D information of scatterers and retrieving the elevation profile. Among the many multi-baseline InSAR techniques, TomoSAR is the only one that strictly reconstructs the full reflectivity along the third dimension elevation. SAR tomography and its differential form (D-TomoSAR) have been extensively developed in last two decades \cite{bib:reigber2000first} \cite{bib:gini2002layover} \cite{bib:lombardini2005differential} \cite{bib:fornaro2005three} \cite{bib:fornaro2009four} \cite{bib:zhu2010very} \cite{bib:ge2018spaceborne} \cite{bib:zhu2018review}. They are excellent approaches for reconstructing the urban area and monitoring the deformation, especially when using high resolution data like TerraSAR-X  \cite{bib:zhu2012demonstration} \cite{bib:zhu2013tomo} or COSMO-Skymed \cite{bib:fornaro2014multilook}. Compare to the classic multi-baseline InSAR algorithms, compressive sensing (CS) based methods  \cite{bib:zhu2010tomographic} \cite{bib:budillon2011three} can obtain extraordinary accuracy for TomoSAR reconstruction and show the super-resolution (SR) power, which is very important for urban areas, since layover is dominant.

Although TanDEM-X bistatic data has many advantages, there is only a limited number of acquisitions available for most areas. For a reliable reconstruction, SAR tomography usually requires fairly large interferometric stacks ($> \textrm{20}$ images), because the variance of the estimates is asymptotically related to the product of SNR and the number of acquisitions. Therefore, it is not appropriate for the micro-stacks, which have limited number of interferograms \cite{bib:zhu2012super}.

\subsection{The proposed framework}
As mentioned above, the accuracy of 3-D reconstruction replies on the product of SNR and the number of measurements $N$. Since the motivation of our work is the large-scale urban mapping, the data we adopted is TanDEM-X stripmap co-registered single look slant range complex (CoSSC), whose resolution is about 3.3 m in azimuth direction and 1.8 m in range direction. The typical number of available interferograms for most areas is 3 to 5 \cite{bib:rizzoli2017generation}. In \cite{bib:zhu2015joint}, the pixels with similar height are grouped for the joint sparsity estimation, which leads to an accurate inversion of TomoSAR using only six interferograms. Although the unprecedented result is obtained, the accurate geometric information is usually not available for most areas. Therefore, the feasible way to keep the required precision of the estimates is to increase the SNR.

Recent works \cite{bib:dhondt2018nonlocal} \cite{bib:ferraioli2018nonlocal} \cite{bib:shi2018non} show that SNR can be dramatically increased by applying non-local filters to the TomoSAR processing for different sensors, such as airborne E-SAR, COSMO-Skymed and TerraSAR-X. In \cite{bib:dhondt2018nonlocal}, different non-local filters have been adopted to improve the estimation of the covariance matrix for distributed scatterers, which leads to a better height estimation for simulated data and airborne SAR data. Ferraioli et al. \cite{bib:ferraioli2018nonlocal} introduced the non-local filter and the total variation regularizer to improve the multi-baseline phase unwrapping process. In \cite{bib:shi2018non}, it is shown that we can achieve a reasonable reconstruction using only seven interferograms and better super-resolution properties when the number of interferograms is relative low.

In this work, we extend the concept of non-local compressive sensing TomoSAR in \cite{bib:shi2018non} \cite{bib:shi2018sar} \cite{bib:shi2019non} and propose a new framework of spaceborne multi-baseline SAR tomography with TanDEM-X bistatic micro-stacks, i.e. 3 to 5 interferograms. The framework includes non-local filtering, spectral estimation, model selection and robust height estimation. Since the different spectral estimators have different estimation accuracy and computational cost, we compared the estimation accuracy of different estimators with micro-stacks.

The demonstration of different TomoSAR inversion methods for a large-scale area has been shown in \cite{bib:wang2014efficient} \cite{bib:shi2018fast} \cite{bib:zhu2013tomo}. Only a few works on the validation of single buildings were reported in \cite{bib:ge2018spaceborne} \cite{bib:zhu2012demonstration} \cite{bib:budillon2017extension}. Therefore, the validation of the specified quality of the TomoSAR result at a larger scale, would be of considerable interests for the scientific and commercial users. We choose Munich city as a test site because of a high quality LiDAR reference available to us, and we propose a complete workflow to compare the TomoSAR point cloud \cite{bib:otepka2013georeferenced} generated by the proposed framework, TanDEM-X DEM product, and LiDAR data.

\subsection{Contribution of this work}
The major contributions of this work are summarized as follows;

\begin{itemize}
\item We make possible a new application of bistatic SAR data for global building height reconstruction.
\item We have pointed out that for pixel-wise multi-master TomoSAR the well-known system equation is no longer valid in the multi-scatterer case.
\item We have developed a framework for tomographic stacks with only 3 - 5 interferograms. A systematic investigation on the estimation accuracy and super-resolution power for the micro-stacks have been carried out, which was never done before.
\item We use 5 TanDEM-X bistatic data to demonstrate the proposed framework. A systematic validation for large-scale TomoSAR reconstruction have been carried out and a method for comparing with other reference data is established. The results are quantitatively compared with LiDAR reference for more than 34,000 buildings.
\end{itemize}

The paper is organized as follows: In section II, the non-local TomoSAR framework is introduced; In section III, the estimation accuracy of TomoSAR with small stacks has been systematically studied; The experiments using real data, is presented in section IV; In section V, the quantitative validation is carried out; Finally, conclusions are given in section V.

\section{Non-Local TomoSAR for Multi-Master InSAR}
\begin{figure}[!ht]
  \centering
  \includegraphics[width=0.5\textwidth]{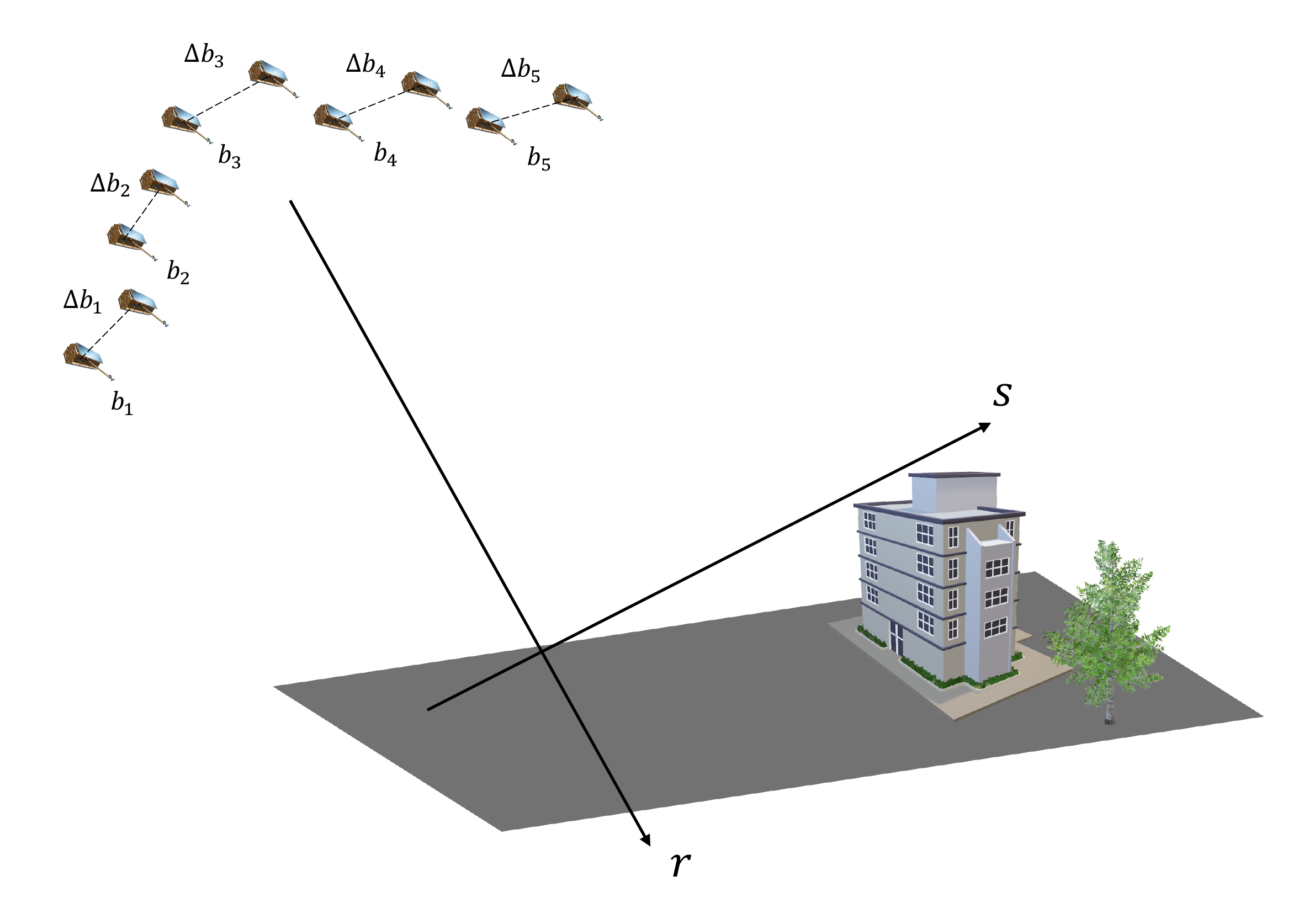}
  \caption{Illustration of multi-master multi-baseline SAR imaging.}
  \label{fig:illu_multi_master}
\end{figure}

\begin{figure*}
  \centering
  \includegraphics[width=1.0\textwidth]{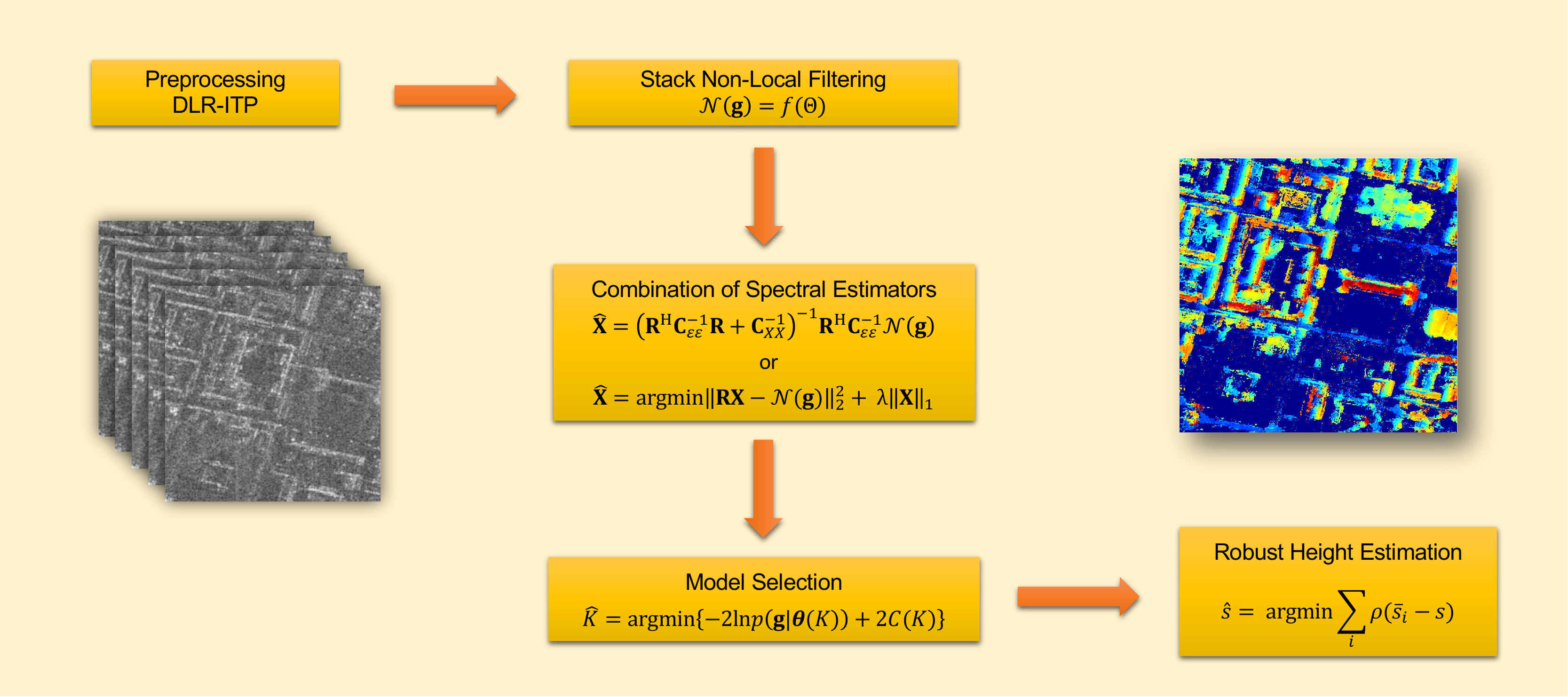}
  \caption{Workflow of non-local TomoSAR framework.}
  \label{fig:workflow_nltomosar}
\end{figure*}

In this section, we introduce the non-local TomoSAR framework for multi-master multi-baseline InSAR configuration. Fig. \ref{fig:illu_multi_master} illustrates multi-master multi-baseline SAR imaging. The framework consists of several steps: (1) non-local filtering; (2) spectral estimation; (3) model selection; (4) robust height estimation. Fig. \ref{fig:workflow_nltomosar} shows the flowchart of the non-local TomoSAR framework.

\subsection{The multi-master TomoSAR imaging model}

For a fixed azimuth-range position, $\gamma(s)$ represents the reflectivity profile along elevation $s$. The measurement $g_n$, i.e. the complex value of the corresponding azimuth-range pixel, in the $n^{th}$ SAR image is then a sample of $\Gamma(k)$ -- the Fourier transform of $\gamma(s)$, where the elevation wavenumber $k$ is a scaled version of the sensor’s position $b_n$ projected on the cross-range-azimuth axis $b||s$ :

\begin{equation}
  g_n = \Gamma(k_n) = \int \gamma(s) \exp(-jk_ns)ds
\end{equation}
with
\begin{equation}
  k_n = -\dfrac{4 \pi b_n}{\lambda r}
\end{equation}
Note that $b_n$ are no baselines, but the positions of the sensor w.r.t. some origin. In case of monostatic multi-temporal data stacks, a \textit{single} master $g_0$ is chosen with $b_0$ and its phase is subtracted from all other acquisitions: $g_ng_0^*/\lvert g_0 \rvert$. This operation renders the phase spatially smooth and is a prerequisite for spatial phase unwrapping, averaging and graph (network) processing. It does not reduce information, since the phase of any (master) acquisition is random. Note that the choice of the point $b=0$ only defines the x-r-s coordinate system. This point need not necessarily be the master track position $b_0$. However, $b_0 = 0$ is a mathematically convenient choice and is assumed in all conventional TomoSAR system model like in \cite{bib:fornaro2003three}. All the equations in \cite{bib:fornaro2003three}, however, are actually independent of this particular choice. Only in the special case $b_0 = 0$ the $b_n$ are identical to baselines.

Here we are dealing with stacks of \textit{bistatic} acquisitions, i.e. with the \textit{multi}-master case. From each of these acquisitions we get a master $g_{n,m} = \Gamma(k_n)$ taken at $b_{master} = b_n$ and a slave $g_{n,s} = \Gamma(k_n+\Delta k_n)$ image taken at $b_{slave} = b_n + \Delta b_n$, where $\Delta b_n$ is the bistatic baseline (which takes the \textit{effective} positions of the transmit-receive phase center into account). If we used a standard, i.e. single-master, TomoSAR inversion algorithm, we would confuse $\Delta b_n$ and $b_n$. In the case of a single scatterer in $\gamma(s)$, this misinterpretation would do no harm, because the Fourier transform of a single point has a constant magnitude and a linear phase. In order to determine the slope of the phase ramp we can take any two samples and divide their phase difference by the difference in wavenumbers (= baseline). This is no longer true for two or more scatterers. The example of two symmetric and equally strong scatterers makes this clear:

\begin{gather}
  \gamma(s) = \delta(s+s_0) + \delta(s-s_0) \nonumber \\
  \updownarrow \\
  \Gamma(k) = 2\cos(s_0k)= 2\cos(2\pi \dfrac{2s_0}{\lambda r}b) \nonumber
\end{gather}

Hence, acquisitions with the same baseline $\Delta b$ are different depending on where the two sensors were located along $b$. Every bistatic acquisition provides three pieces of information: the two magnitudes $|\Gamma(k_n)|$ and $|\Gamma(k_n+\Delta k_n)|$ as well as the phase difference $\angle \Gamma(k_n+\Delta k_n)\Gamma^*(k_n)$, i.e. we must normalize the phase by the respective master in every acquisition, in order to become unaffected by deformation and atmospheric delay. Spectral estimation based conventional TomoSAR inversion algorithms, however, require complex spectral samples at several wavenumbers, phase-normalized to a \textit{single} master phase. In a current parallel work by one of the authors \cite{bib:ge2019single} it is shown that pixel-wise TomoSAR using multi-master acquisitions is a non-convex hard to solve problem.

This is true for \textit{pixel-wise} tomographic inversion or for point scatterers. The situation becomes different, though, once we talk about averages of pixels, i.e. estimates of expectation values. Let us assume Gaussian distributed scattering with a backscatter coefficient along elevation of

\begin{equation}
  \sigma_0(s) = E \left\{ |\gamma(s)|^2 \right\}
\end{equation}

Assuming further that $\gamma(s)$ is white, its power spectral density is stationary and is the autocorrelation function of $\Gamma(k)$, i.e. the Fourier transform of $\sigma_0(s)$ as a function of the baseline wavenumber $\Delta k$ :

\begin{equation}
  E \left\{ \Gamma(k_n+\Delta k_n) \Gamma^*(k)\right\} = \int \sigma_0(s)\exp(-j\Delta k_n s)ds
\end{equation}

Instead of sampling the Fourier spectrum we sample its autocorrelation function by the bistatic data stack. Since this relationship is \textit{independent} of $k \propto b$ because of stationarity, it makes no difference, where the two acquisitions have been taken, only their baseline $\Delta b_n$ counts. In other words we can use standard TomoSAR inversion algorithms in this case.

In this paper we use nonlocal filtering to improve SNR for micro-stacks. These filters perform ensemble averages with number of looks in the order of tens to hundreds. Hence, we tend to the assumption that we work with reasonably good estimates of $E \left\{ \Gamma(k_n+\Delta k_n) \Gamma^*(k)\right\}$ and can use the bistatic \textit{interferograms} for TomoSAR reconstruction.

By introducing a noise $\boldsymbol{\varepsilon}$, the matrix notation of TomoSAR model can be formulated as:
\begin{equation}
\mathbf{g}=\mathbf{R}\mathbf{X}+\boldsymbol{\varepsilon}
\label{equ:tomosar_basic}
\end{equation}
where $\mathbf{g} = [g_1, g_2, ..., g_n]^{\mathrm{T}}$ is vector notation of the complex-valued measurement with dimension $N \times 1$, and $\mathbf{X} \sim \sigma_0(s_l) = E\{|\gamma(s_l)|^2\}$ is the expectation value of reflectivity profile along elevation uniformly sampled at $s_l (l=1,2,...,L)$. $\mathbf{R}$ is a sensing matrix with the dimension $N \times L$, where $R_{nl} = \exp(-j\Delta k_ns_l)$.

\subsection{Non-Local Procedure}
Since we have only limited number of acquisitions for large-scale area, the SNR need to be dramatically increased in order to obtain the required accuracy. As shown in \cite{bib:shi2018non}, non-local procedure is an efficient way to increase the SNR of interferograms without notable resolution distortion. The idea of patch-wise non-local means considers all the pixels $s$ in the search window, when the patch with the central pixel $s$ is similar to the patch with central pixel $c$, the value of $s$ is selected for calculating the value of pixel $c$. The value of pixel $c$ is estimated by using a weighted maximum likelihood estimator (WMLE).
\begin{equation}
\hat{\boldsymbol{\Theta}}_c = \mathrm{argmax} \sum_s \mathbf{w}(i_s, j_s) \log p(\mathbf{g}_s|\boldsymbol{\Theta})
\end{equation}
where weights $\mathbf{w}(i_s, j_s)$ can be calculated by using patch-wise similarity mesurement \cite{bib:shi2018non}. Assuming that we have two expressions $\mathbf{g} = (I_1, I_2, \phi)$ and $\boldsymbol{\Theta} = (\psi, \mu, \sigma^2)$, where $\mathbf{g}$ denotes the complex-valued measurement. $I_1$ and $I_2$ are the instensity of two SAR images. $\phi$ is the interferometric phase. $\boldsymbol{\Theta}$ is the true value of the parameters, where $\psi$ is the noise-free interferometric phase, $\mu$ is the coherence magnitude, and $\sigma^2$ is the variance. The likelihood function $p \left( \mathbf{g}_s | \boldsymbol{\Theta} \right) = p \left(I_{1,s}, I_{2,s}, \phi_{s} | \psi, \mu, \sigma^2 \right)$ is adopted from \cite{bib:goodman2007speckle} with following formualtion:
\begin{multline}
  p(I_1, I_2, \phi | \psi, \mu, \sigma^2) = \dfrac{1}{16\pi^2 \sigma^4(1-\mu^2)}  \\ \times \exp \left[-\dfrac{I_1 + I_2 - 2 \sqrt{I_1 I_2}\mu \cos(\phi - \psi)}{2\sigma^2(1-\mu^2)} \right]
\end{multline}

$\mathcal{N}(.)$ denotes the non-local estimator, where $\mathcal{N}(\mathbf{g}) = f( \hat{\boldsymbol{\Theta}})$. $\hat{\boldsymbol{\Theta}} = (\hat{\psi}, \hat{\mu}, \hat{\sigma^2})$ represents the parameters being estimated, where $\hat{\psi}$ is the estimated interferometric phase, $\hat{\mu}$ stands for the coherence magnitude, and $\hat{\sigma^2}$ stands for the variance.  $f( \hat{\boldsymbol{\Theta}})$ is the maximum likelihood estimator and the estimated parameters can be formulated as
\begin{eqnarray}
  \hat{\psi}  &=& -\arg \left(\sum_s \mathbf{w}_s \mathbf{g}_{1,s} \mathbf{g}_{2,s}^{*} \right) \\
  \hat{\mu}   &=& \dfrac{2 \sum_s \mathbf{w}_s |\mathbf{g}_{1,s}| |\mathbf{g}_{2,s}|}{\sum_s \mathbf{w}_s \left(|\mathbf{g}_{1,s}|^2 + |\mathbf{g}_{2,s}|^2 \right)} \\
  \hat{\sigma^2} &=& \dfrac{\sum_s \mathbf{w}_s \left(|\mathbf{g}_{1,s}|^2 + |\mathbf{g}_{2,s}|^2 \right)}{4 \sum_s \mathbf{w}_s}
\end{eqnarray}

The patch size and search window size are set to be 7 $\times$ 7 and 21 $\times$ 21 according to experimental study, which is also reported by other works \cite{bib:deledalle2011nl} \cite{bib:zhu2018potential}. Each pixel represents 2.17 m in azimuth and 1.36 m in range.

\begin{figure*}
  \centering
  \subfloat[]{\includegraphics[width=0.48\textwidth]{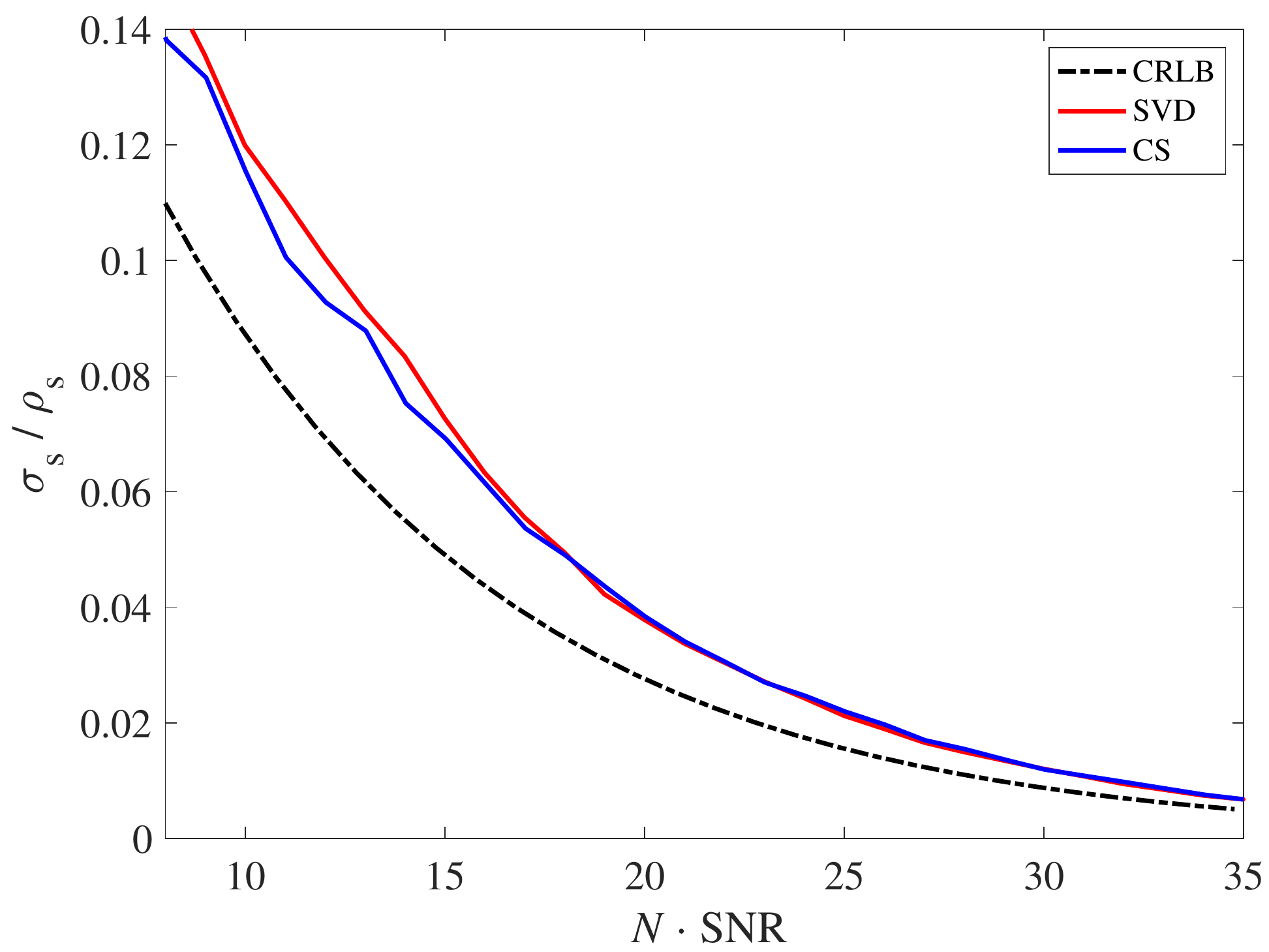}}
  \subfloat[]{\includegraphics[width=0.48\textwidth]{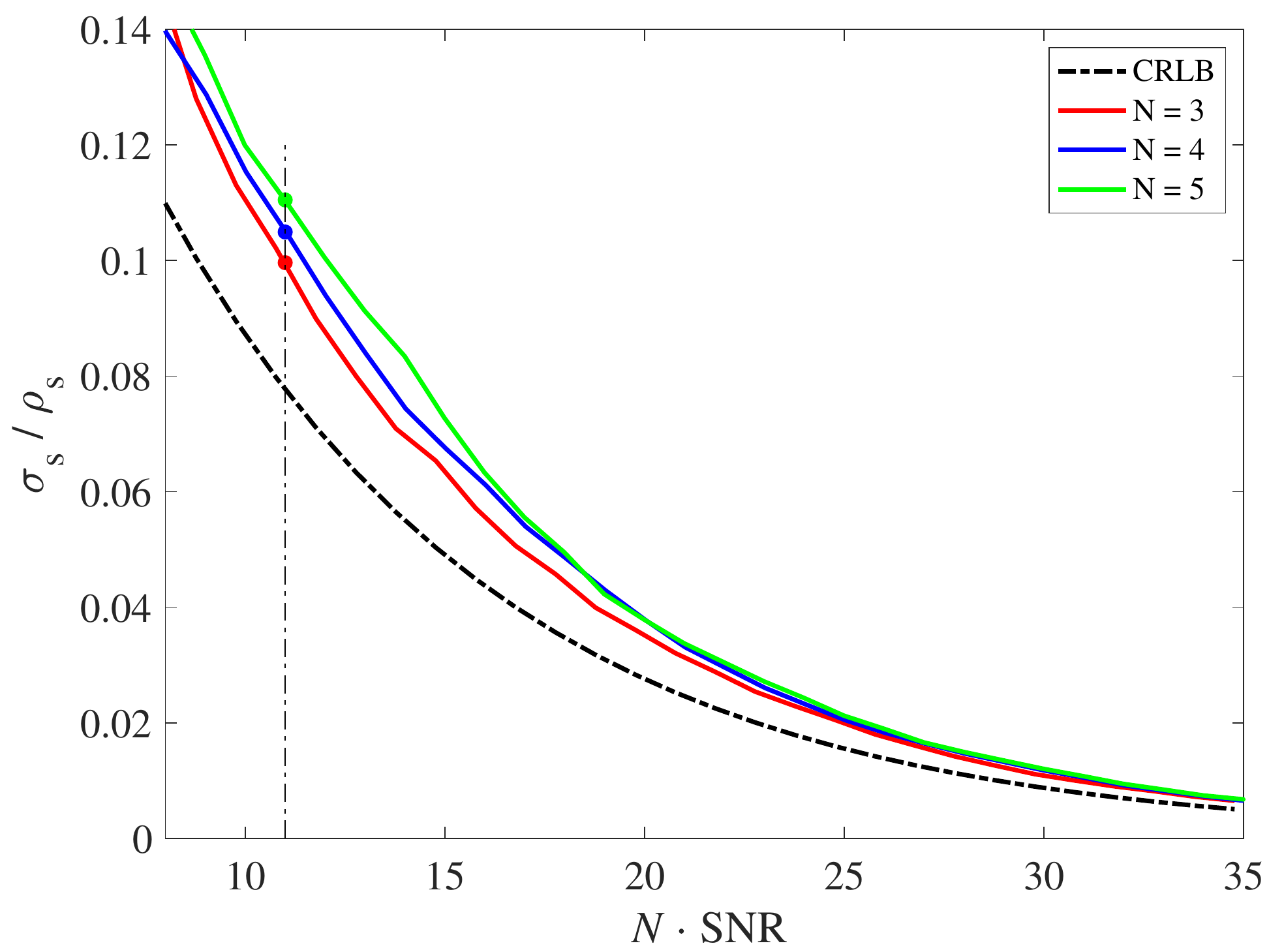}}
  \caption{Monte Carlo simulations of single scatterer with SNR in [0 30] (dB). X-axis presents $N \cdot \mathrm{SNR}$ in dB. Y-axis is the normalized CRLB $\sigma_s / \rho_s$.  (a) Comparison of CRLB with different spectral estimators with five acquisitions. SVD (red solid line), CS (blue solid line), CLRB (black dash-dotted line) (b) Comparison of CRLB using SVD with three to five acquisitions. $N=3$ (red solid line), $N=4$ (blue solid line), $N=5$ (green solid line), CLRB (black dash-dotted line). The vertical black dash-dotted line indicates the estimation accuracy for $N \cdot \textrm{SNR} =  11$ dB. The red, blue and green markers represent $N = 3,4,5$, respectively. }
  \label{fig:simulation_single_scatterer}
\end{figure*}

\subsection{Spectral Estimation}
After the non-local procedure, spectral estimation is applied. The most relevant spectral estimation algorithms, including singular value decomposition (SVD) \cite{bib:fornaro2005three} \cite{bib:zhu2010very}, compressive sensing (CS) are introduced in the following.
\begin{itemize}
\item SVD:
\begin{equation}
  \hat{\mathbf{X}} = \left( \mathbf{R}^{\textrm{H}} \mathbf{C}_{\varepsilon\varepsilon}^{-1} \mathbf{R} +  \mathbf{C}_{XX}^{-1} \right)^{-1} \mathbf{R}^{\textrm{H}} \mathbf{C}_{\varepsilon\varepsilon}^{-1} \mathcal{N}(\mathbf{g})
\end{equation}

\item CS:
\begin{equation}
\hat{\mathbf{X}} = \arg \min_{\mathbf{X}} \{ \Vert \mathbf{R}\mathbf{X} - \mathcal{N}(\mathbf{g}) \Vert^2_2 + \lambda \Vert \mathbf{X} \Vert_1 \}
\label{equ:opt_nll1lsp}
\end{equation}
\end{itemize}
where $\mathbf{C}_{\varepsilon\varepsilon}$ is the noise covariance matrix, which is defined as:
\begin{equation}
  \mathbf{C}_{\varepsilon\varepsilon} = \left( \mathbf{g} - \mathbf{R}\mathbf{X} \right) \cdot \left( \mathbf{g} - \mathbf{R}\mathbf{X} \right)^\mathrm{H}
\end{equation}

Under the assumption that the model errors are circular Gaussian distributed with zero mean, the noise covariance matrix is formulated as $\mathbf{C}_{\varepsilon\varepsilon} = |\sigma_{\varepsilon}|^2 \mathbf{I}$ and $|\sigma_{\varepsilon}|^2$ is the noise power level. $\mathbf{C}_{XX}$ is the covariance matrix of the prior, if it is assumed to be white, i.e. $\mathbf{C}_{XX} = \mathbf{I}$.

The choice of different combinations of spectral estimators depends on the required accuracy, the computational time and others. We follow the procedure proposed in \cite{bib:wang2014efficient}. It consists of three parts: (1) an efficient low-order spectral estimation; (2) the discrimination of the number of scatterers; (3) an accurate high-order spectral estimation. The elevation profile is first estimated by an efficient low-order spectral estimator in order to discriminate the number of scatterers in one resolution cell. Then, CS-based approach is adopted for the pixel which has multiple scatterers. This method decreases the amount of pixels that need the $L_1$ minimization, which leads to reduce the computational cost. Furthermore, the rest of pixels can be efficiently solved by randomized blockwise proximal gradient method \cite{bib:shi2018fast}.

\subsection{Model Selection}
The abovementioned spectral estimators retrieve a nonparametric reflectivity profile. Since our data is in urban area, we assume only a few dominant scatterers exist along the reflectivity profile. Therefore, the number of scatterers $\hat{K}$ is estimated by a model order selection algortihm as well as their elevation in one azimuth-range pixel \cite{bib:zhu2010very}. The estimator can be expressed as follows.

\begin{equation}
  \hat{K} = \arg \min_{K} \left\{ -2 \ln p \left( \mathbf{g}| \boldsymbol{\theta}\right) + 2 C(K) \right\}
\end{equation}
where $C(k)$ is a model complexity penalty term which avoids more complicated model overfitting the observed data. The classical penalized likelihood criteria are the Bayesian information criterion (BIC), the Akaike information criterion (AIC), and the minimum description length (MDL) principle \cite{bib:lombardini2005model}.

As mentioned in \cite{bib:zhu2010very}, the criteria of model order selection has to be chosen according to the experiments for the particular situation, because it is difficult to remove the bias of the selection.

\subsection{Robust Height Estimation}
To tackle the possible remaining outliers in the height estimates, the final height will be fused from the result of multiple neighbouring pixels as a post-processing. But instead of simple averaging, the height will be adjusted robustly using an \textit{M-estimator}. Instead of minimizing the sum of squared residuals in averaging, M-estimator minimizes the sum of a customized function $\rho\left(\centerdot\right)$ of the residuals:
\begin{equation}
\tilde{s}=\mathop{\arg}\underset{\textit{s}}{\mathop{\min}}\sum\limits_{i}{\rho\left(\hat{s}_i- s\right)},
\label{eq:M_estimator}
\end{equation}
where $\hat{s}_i$ is the elevation estimates of the $i$th neighbouring pixel. It is shown that the close-formed solution of Eq. (\ref{eq:M_estimator}) is simply a weighted averaging of the heights of the neighbouring pixels \cite{bib:wang2016robust}. The weighting function can be expressed as follows, if the derivative of $\rho\left(x\right)$ exists.
\begin{equation}
w\left(x\right)=\frac{\partial\rho\left(x\right)}{x\partial x}
\end{equation}

The robust estimated height can be written as follows:
\begin{equation}
  \tilde{h} = \dfrac{\sum \limits_{i} w(x_i) \cdot \hat{h}}{\sum \limits_{i} w(x_i)}
\end{equation}
where $\hat{h} =\hat{s} \cdot \sin\theta$, and $\theta$ is incident angle. The choice of the weighting function depends on the distribution of the heights. Without prior knowledge of the distribution, promising robust weighting functions are Tukey's biweight or t-distributed weighting \cite{bib:wang2016robust}. For instance, the formulation of Tukey's biweight loss function can be written as:
\begin{equation}
  \rho(x) = \left\{
  \begin{array}{ccl}
-\frac{\left(c_r^2-x^2\right)^3}{6c_r^4} + \frac{c_r^2}{6} & & {|x| < c_r}\\
\frac{c_r^2}{6} & & {\mathrm{elsewhere}}
\end{array}
\right.
\end{equation}
and the weighting function can be formulated as:
\begin{equation}
  w(x) = \left\{
  \begin{array}{ccl}
  1 - \frac{x^4}{c_r^4} - \frac{2x^2}{c_r^2} & & {|x| < c_r}\\
  0 & & {\mathrm{elsewhere}}
  \end{array}
  \right.
\end{equation}

\section{Estimation accuracy of TomoSAR with small stacks}
This section will discuss the theoretical 3-D reconstruction accuracy of a micro-stack with 3-5 interferograms. The estimation accuracy of TomoSAR has been systematically investigated. It is exhaustively shown in \cite{bib:zhu2012super} that the elevation estimation accuracy and SR power depend asymptotically on the multiplication $N \cdot \textrm{SNR}$. In this section, we investigate the estimation accuracy of TomoSAR with the extremely small number of interferograms, which is 3 to 5.

\begin{figure*}
  \centering
  \subfloat[]{\includegraphics[width=0.33\textwidth]{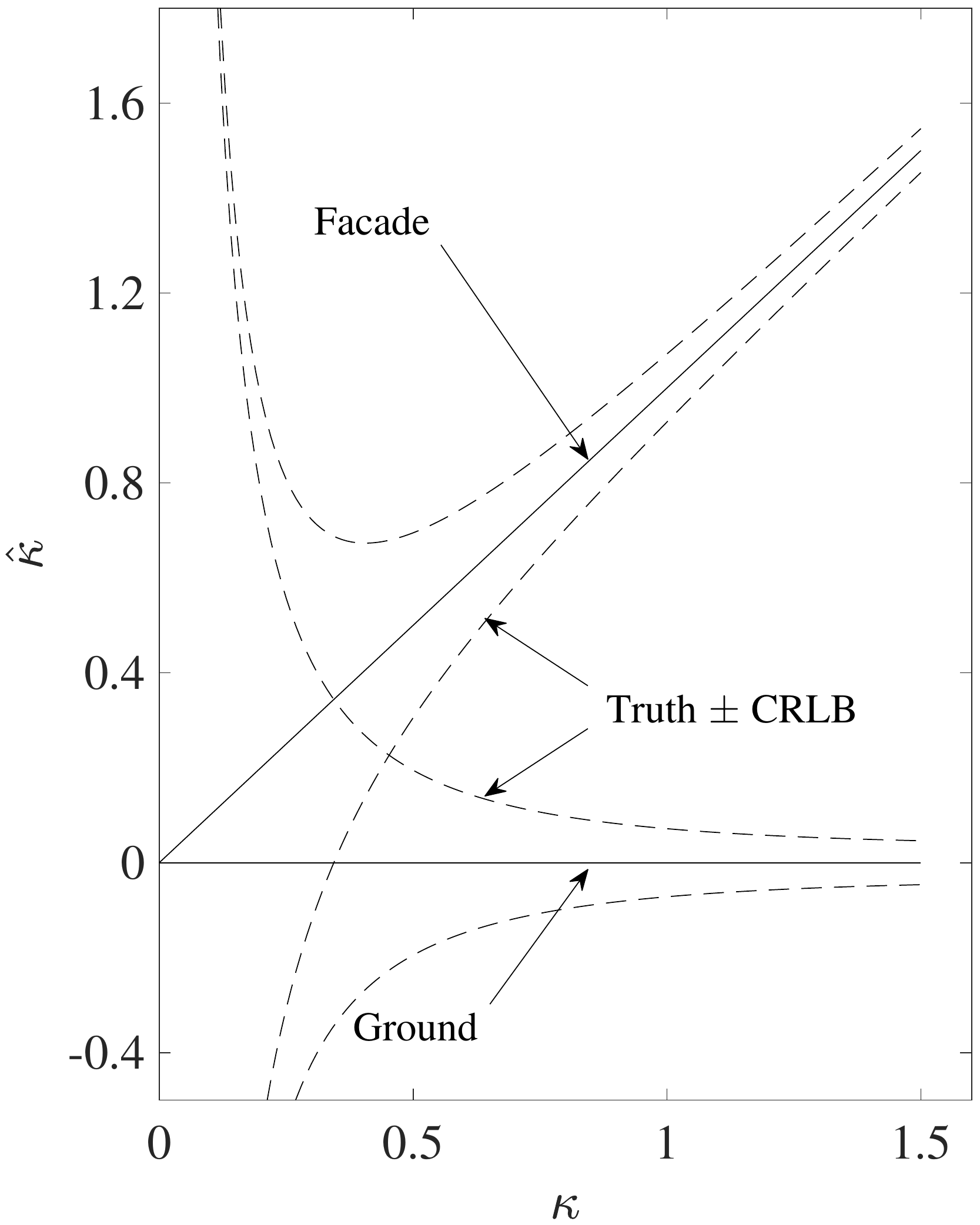}}
  \subfloat[]{\includegraphics[width=0.33\textwidth]{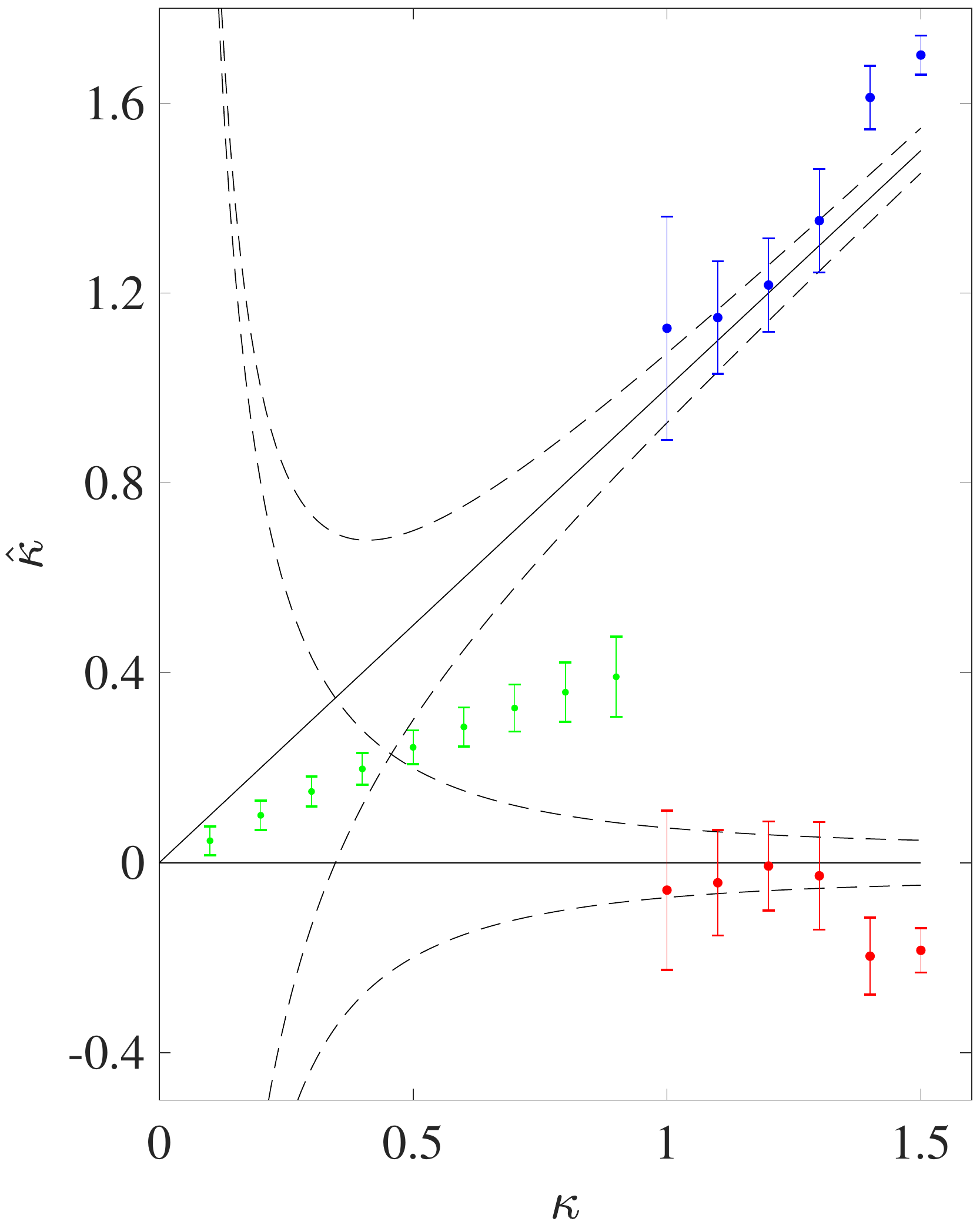}}
  \subfloat[]{\includegraphics[width=0.33\textwidth]{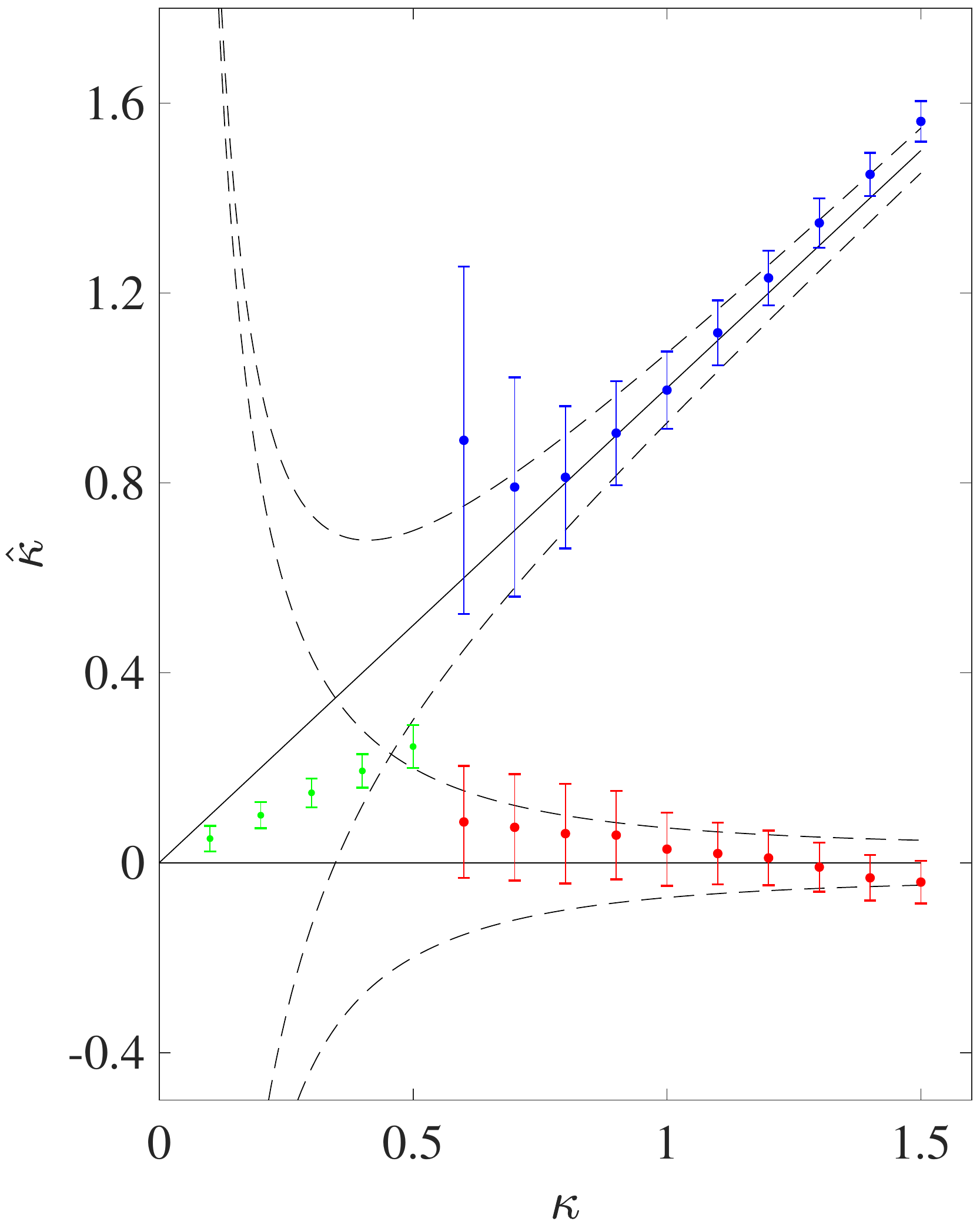}}
  \caption{Monte Carlo simulations of double scatterer with different normalized distances: $\kappa \in [0.1, 1.5]$ and $\textrm{SNR} = 10$ dB. X-axis represents normalized true distance $\kappa$ of simulated facade and ground. Y-axis is normalized estimated distance $\hat{\kappa}$ of simulated facade and ground. The blue dot marker denotes the estimated location of facade and the error bar indicates the standard deviation of the estimates, whereas the red dot marker represents the estimated location of ground. The green dot suggests that detection rate of double scatterers is below 5\% and denotes the estimated result of single scatterer. (a) Illustration (b) SVD (c) CS. }
  \label{fig:simulation_double_scatterers}
\end{figure*}

\subsection{The Lower Bound for Micro-stacks}
In the case of pixel-wise TomoSAR inversion, i.e. without spatial averaging, each of our $N$ bistatic pairs contain three pieces of information, as mentioned before. If we want to reconstruct elevation profiles containing $M$ discrete scatterers we need to infer $3M$ parameters, i.e. elevation, magnitude and phase for each scatterer. Hence an absolute lower bound of the micro-stack size is $N \geq M$.

Distributed scatterers, on the other hand, are characterized by only two parameters each: elevation and backscatter coefficient. Likewise each interferogram provides only two parameters, magnitude and phase (difference). Since our goal is 3-D reconstruction based on bistatic data, we disregard motion-induced phase here. Hence, also in this case the absolute lower limit is $N \geq M$. This limit is only a necessary condition, however not sufficient from the robustness point of view, because of ambiguities in the inversion cost functions.

For 3-D urban mapping the single and double scattering cases are the dominant ones. We investigate the cases $N = 3 - 5$ in this paper, because these are close to the mentioned limits and are relevant for TanDEM-X.

\subsection{CRLB}
It is demonstrated in \cite{bib:zhu2010very} that the Cramer-Rao lower bound (CRLB) of the elevation estimates for single scatterer can be expressed as:
\begin{equation}
  \sigma_{s} = \dfrac{\lambda r}{4 \pi \cdot \sigma_b \cdot \sqrt{2 \cdot \textrm{SNR} \cdot N}}
  \label{eq:crlb}
\end{equation}
where $\sigma_b$ is the standard deviation of the baseline distribution. $N$ is the number of interferograms, and SNR is the signal-to-noise ratio.

For the double scatterers' case, the CRLB can be written as:
\begin{equation}
  \sigma_{s_q} = c_0 \cdot \sigma_{s_q,0}
\end{equation}
where $\sigma_{s_q,0}$ represents the CRLB on the elevation estimation of the $q$th scatterer without the interference with the others. $c_0$ is the correction factor of the interference for the scatterers, which are closely located \cite{bib:zhu2012super}. It is nearly free from $N$ and SNR, which can be written as:
\begin{equation}
  c_0 = \max \left \{ \sqrt{\dfrac{40\kappa^{-2}(1-\kappa/3)}{9-6(3-2\kappa) \cos(2 \Delta \varphi)+(3-2\kappa)^2}}, 1 \right \}
\end{equation}
where $\Delta \varphi$ is the phase difference of the two scatterers. $\kappa$ is the normalized distance between two scatterers (defined in next section). Since $\Delta \varphi$ is a random variable, the approximated formulation of $c_0$ can be calculated by integrating the variances over $\Delta \varphi$.
\begin{equation}
  c_0 = \max \left \{ 2.57(\kappa^{-1.5} - 0.11)^2 + 0.62, 1 \right \}
\end{equation}

A note on the baseline distribution is worth mentioning: all the Eqs. (21)-(24) are satisfied with large stacks. In the micro-stacks with only 3 - 5 acquisitions, the baseline distribution may be unfavorable, even if the baseline spread $\sigma_b$ is acceptable. For example, if two baselines were very similar, the information content would be reduced. It is therefore desirable to have the baselines possibly statistically uniformly distributed.

\begin{figure*}
  \centering
  \subfloat[]{\includegraphics[width=0.95\textwidth]{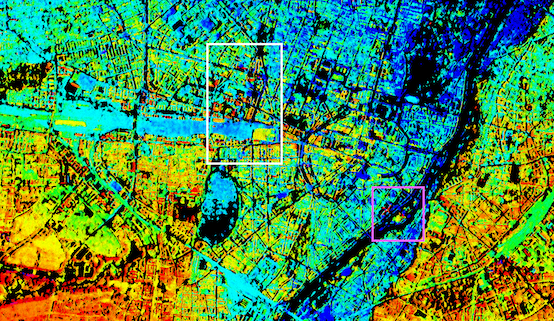}}
  \hfil
  \subfloat[]{\includegraphics[width=0.95\textwidth]{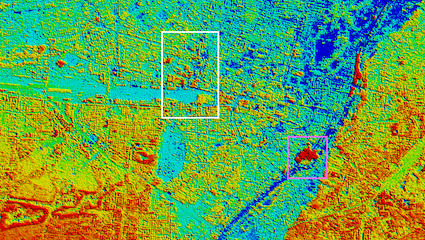}}
  \caption{Visual comparison of NL-TomoSAR point clouds and TanDEM-X DEM over Munich, Germany. Color code: 565 m (blue) – 596 m (red), scene size: 15 km $\times$ 9 km, north = top. The voilet bounding box indicates the region of interest (ROI) over the area of European bureau of patent and the white bounding box indicates the ROI near Munich central station. (a) Point Clouds generated by NL-TomoSAR with five interferograms. (b) TanDEM-X DEM.}
  \label{fig:comp_dem_tomosar_large}
\end{figure*}

\subsection{Monte Carlo Simulations}
In this section, we compare different spectral estimators using simulated data. Two cases were carried out. The first case considers only a single scatterer in the interest of exploring the effect of $N$ and SNR on the estimation accuracy for micro-stacks and the performance of different estimators. The second case considers double scatterers to investigate the estimation accuracy and the super-resolution power for different estimators. The inherent (Rayleigh) elevation resolution $\rho_s$ is inversely proportional to the maximal elevation aperture $\Delta b$ \cite{bib:zhu2012super}.

\begin{equation}
\rho_s = \dfrac{\lambda r}{2\Delta b}
\end{equation}
The normalized distance is defined as
\begin{equation}
\kappa = \dfrac{s}{\rho_s}
\end{equation}

For the first test case, only one scatterer is placed at $s = 0$, and the SNR is in the range between 0 and 30 dB. For each $N \cdot \textrm{SNR}$ value, 100 different baseline distributions were generated. We carried out a Monte Carlo simulation for each baseline distribution with 10,000 realizations. Afterwards, the CRLB was evaluated by averaging the value of 100 different baselines. Fig. \ref{fig:simulation_single_scatterer} (a) shows a performance comparison between SVD, and CS on simulated data with five acquisitions for a single scatterer. X-axis presents $N \cdot \mathrm{SNR}$ in dB. Y-axis is the normalized CRLB $\sigma_s / \rho_s$. As one can see, both approaches have similar estimation accuracy. They are asymptotically towards the CRLB and collapse it when $N \cdot \mathrm{SNR}$ is large. More interestingly is when $N \cdot \textrm{SNR}$ is fixed, leaving $N$ as a variable. Fig. \ref{fig:simulation_single_scatterer} (b) presents the estimation accuracy of SVD with $N = 3,4,5$. It shows that the accuracy when $N=3$ is the smallest. This indicates that SNR carries more weight than $N$ on the estimation accuracy when $N$ is very small.

With the 3-D reconstruction accuracy of single scatterer clearly analyzed, we switch to the double scatterers case. In the simulation, the elevation of one scatterer is fixed at 0. the normalized elevation of the other scatterer is increased from 0.1 to 1.5, in order to mimic the layover of a ground layer and a facade layer. The number of acquisitions is set to $N = 3-5$, same as the first simulation. SNR is set to be 10 dB, since the SNR of TanDEM-X bistatic data is usually higher than this value in urban area \cite{bib:zhu2012sparse}.

The Monte Carlo simulation result is shown in Fig. \ref{fig:simulation_double_scatterers}. The x-axis represents the true normalized elevation distance $\kappa$ of the simulated facade and ground layers. Y-axis is the estimated normalized elevation distance $\hat{\kappa}$ of the simulated facade and ground layers. The two solid lines in the Fig. \ref{fig:simulation_double_scatterers} (a) (b) (c) represent the true position of the building facade and the ground, respectively. The dashed lines imply the true position plus and minus the CRLB. The blue bar and dot imply the standard deviation and the mean of the estimated elevation of the facade scatterers, whereas the red ones represent those of the ground scatterers. The green dot indicates that the detection rate of double scatterers is below 5\% and denotes the estimated result of the single scatterer. Fig. \ref{fig:simulation_double_scatterers} (b) (c) show the estimated results by SVD and CS, respectively.

As one can see in Fig. \ref{fig:simulation_double_scatterers} (b) (c), the result of SVD has larger bias and slightly bigger standard deviation than CS. Note that, comparing to SVD, CS can give the better result, not only the accuracy of the estimation but also the super-resolution power. As one can see that the SVD has scarcely no super-resolution power, which can only distinguish double scatterers tile one Rayleigh resolution $\rho_s$. In contrast, CS can achieve until 0.6 $\rho_s$.

\section{Practical Demonstration}
\subsection{Data Description}
We make use of a stack of five co-registered TanDEM-X bistatic interferograms to evaluate the proposed algorithm. The dataset is over Munich, Germany, whose slant range resolution is 1.8 m and the azimuth resolution is 3.3 m. The images were acquired from July 2016 to April 2017. The most pertinent parameters of a TanDEM-X bistatic stripe map acquisition of Munich are listed in Table \ref{tab:realData_char} and Table \ref{tab:realData_char1}. All the preprocessing steps, like deramping, are standard that are known from bistatic forest tomography. For interested readers, please refer to \cite{bib:reigber2000first}.
\begin{table}[!ht]
\begin{center}
\caption{Parameters of Tandem-X Stripe Map Acquisition of Munich}
\label{tab:realData_char}
\begin{tabular}{lll}
\toprule
Name &  Symbol & Value\\
\midrule
 Distance from the scene center & $r$ & 698 km \\
\rule{0pt}{4ex}Wavelength & $\lambda$ & 3.1 cm\\
\rule{0pt}{4ex}Incidence angle at scene center & $\theta$ & $50.4^{\circ}$\\
\rule{0pt}{4ex}Maximal elevation aperture & $\Delta b$ & 187.18 m\\
\rule{0pt}{4ex}Number of interferograms & $N$ & 5\\
\bottomrule
\end{tabular}
\end{center}
\end{table}

\begin{table}[!ht]
\begin{center}
\caption{Detailed Information of Tandem-X Stripe Map Acquisition for used Dataset}
\label{tab:realData_char1}
\begin{tabular}{cccc}
\toprule
No. &  Date & Baseline [m] & Height Ambiguity [m/cycle]\\
\midrule
 1 & 2016-07-25 & 184.40 & 50.30 \\
 2 & 2016-09-07 & 171.92 & 54.01 \\
 3 & 2017-02-19 & 32.30  & 286.03 \\
 4 & 2017-04-26 & -2.78  & -8710.99 \\
 5 & 2017-07-01 & 9.30 & 1073.03 \\
\bottomrule
\end{tabular}
\end{center}
\end{table}

\subsection{Visual Comparison with TanDEM-X raw DEM}
In this work, the TanDEM-X raw DEM is adopted for visual comparison with TomoSAR point clouds of the test area, which is formed by two TanDEM-X bistatic acquisitions using the Integrated TanDEM-X Processor (ITP).

\begin{figure}[!ht]
  \centering
  \subfloat[]{\includegraphics[width=0.24\textwidth]{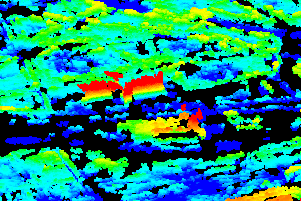}}
  \hfil
  \subfloat[]{\includegraphics[width=0.24\textwidth]{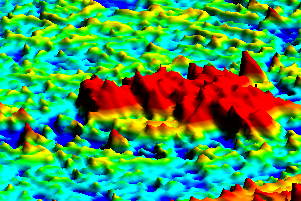}}
  \caption{Visual comparison of NL-TomoSAR point clouds and TanDEM-X DEM, close-up 3-D view over the area of European bureau of patent. (a) TomoSAR point clouds. (b) TanDEM-X DEM.}
  \label{fig:comp_dem_tomosar_patentamt}
\end{figure}

\begin{figure}[!ht]
  \centering
  \subfloat[]{\includegraphics[width=0.5\textwidth]{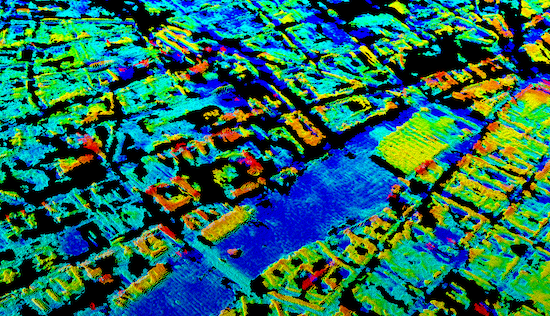}}
  \hfil
  \subfloat[]{\includegraphics[width=0.5\textwidth]{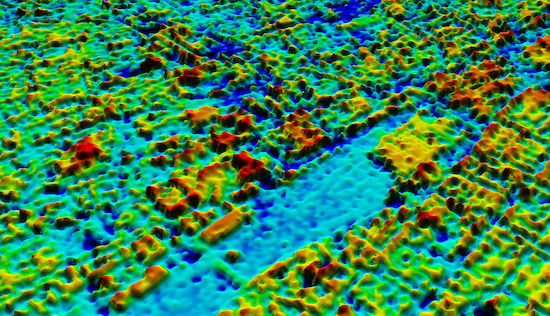}}
  \caption{Visual comparison of NL-TomoSAR point clouds and TanDEM-X DEM, close-up 3-D view over the area of Munich central station. (a) TomoSAR point clouds. (b) TanDEM-X DEM.}
  \label{fig:comp_dem_tomosar_small}
\end{figure}

\begin{figure*}
  \centering
  \subfloat[]{\includegraphics[width=0.3\textwidth]{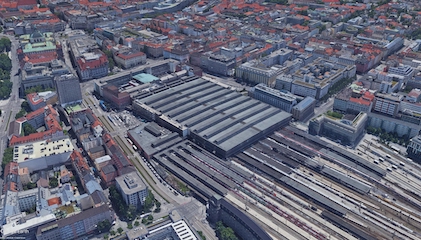}}
  \hfil
  \subfloat[]{\includegraphics[width=0.3\textwidth]{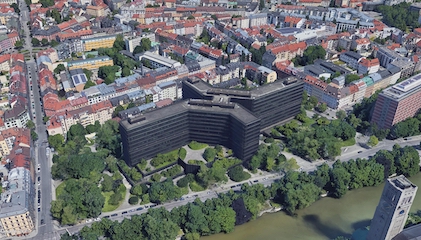}}
  \hfil
  \subfloat[]{\includegraphics[width=0.3\textwidth]{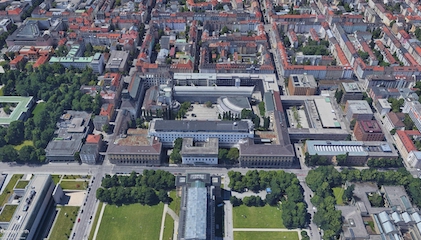}}
  \vfil
  \subfloat[]{\includegraphics[width=0.3\textwidth]{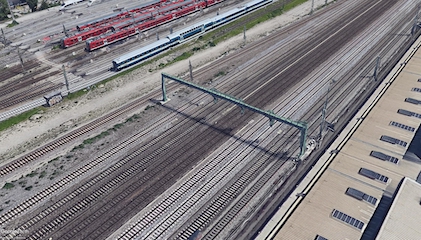}}
  \hfil
  \subfloat[]{\includegraphics[width=0.3\textwidth]{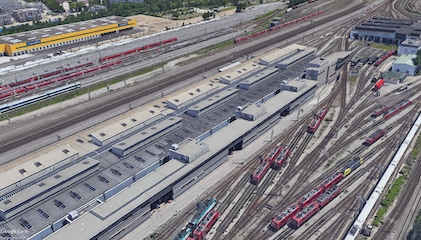}}
  \hfil
  \subfloat[]{\includegraphics[width=0.3\textwidth]{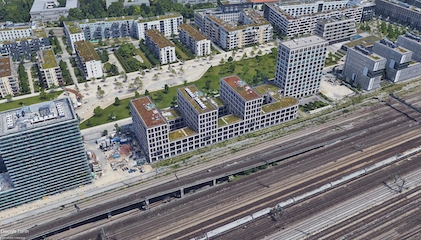}}
  \vfil
  \subfloat[]{\includegraphics[width=0.3\textwidth]{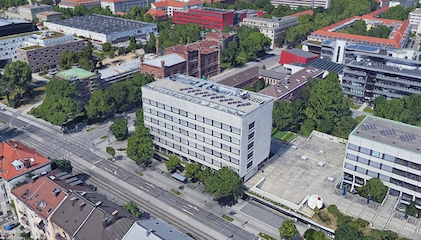}}
  \hfil
  \subfloat[]{\includegraphics[width=0.3\textwidth]{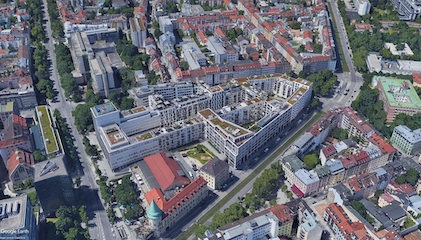}}
  \hfil
  \subfloat[]{\includegraphics[width=0.3\textwidth]{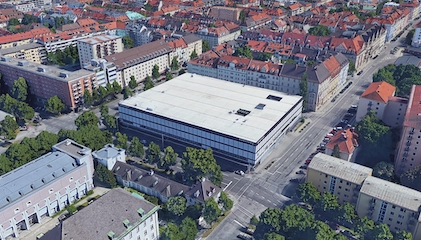}}
  \caption{ Optical images of nine test sites for quantitative comparison of NL-TomoSAR point clouds and TanDEM-X DEM. (a) Munich central station. (b) European bureau of patent. (c) Technical University of Munich. (d) Railway signal light stand. (e) Train repair garage. (f) Residential building between two bridges. (g) Munich University of Applied Sciences. (h) Residential building near Lowenbrau beer company. (i) Karstadt (shopping mall).}
  \label{fig:optical_comp9}
\end{figure*}

A top view of the reconstructed point cloud of TomoSAR is shown in Fig. \ref{fig:comp_dem_tomosar_large} (a). The black regions in the figure is where the pixels are not coherent. The corresponding area of TanDEM-X raw DEM is presented in Fig. \ref{fig:comp_dem_tomosar_large} (b) as a comparison. It is clear that the result of TomoSAR point cloud preserves more detailed building structures. The road layer is also better represented in the TomoSAR result as well. In Fig. \ref{fig:comp_dem_tomosar_large} (b), the flat ground surface are well reconstructed. But when it comes to complex or high-rise buildings, their accuracy is compromised. For instance, the building of European bureau of the patent in the bottom right (red color) along the Isar river. A close view to this building can be seen in Fig. \ref{fig:comp_dem_tomosar_patentamt}. Due to the complex building structure, as well as the multilooking processing, the TanDEM-X raw DEM merges several buildings together and exhibits lower accuracy on the height of the buildings.

As another example, Fig. \ref{fig:comp_dem_tomosar_small} shows the visual comparison over the area around Munich central station. It is clear that NL-TomoSAR result can show more detailed structures, such as the bridge, the central station, and roads.

\section{Quantitative Validation}
In this section, we have quantitatively compared the TomoSAR point clouds with TanDEM-X raw DEM, as well as a much more precise LiDAR reference. The LiDAR dataset of Munich is provided by Bavarian State Office for Survey and Geoinformation with ten centimeter accuracy \cite{bib:lidar2017munich}.

Since the TomoSAR point cloud is with respect to a reference point that was chosen during the TomoSAR processing, its location is not with respect to a geo-coordinate system. We coregistrated the point cloud of TomoSAR with the DEM and the LiDAR point cloud. In addition, in order to compare point clouds with DEM, we rasterize the two point clouds. These preparing steps are briefly explained in this section.

\subsection{Geocoding}
Since the result of TomoSAR inversion is a 3-D point cloud in the range-azimuth coordinate, the first step is to transform the result to Universal Transverse Mercator (UTM) coordinate with the Range-Doppler approach \cite{bib:schwabisch1998fast}.

\begin{table*}
\centering
\caption{Statistics of quantitative comparison of nine test structures. First column shows the number of each structure. Second column implies the source of each result, i.e., t (tomosar), l (lidar), d (dem). Third and Fourth columns present the statistics (min, max, mean and standard deviation) of sample points at top layer and bottom layer. Fifth column demonstrates the relative height of each structure, which is calculated by using the mean value of top layer minus the mean value of bottom layer. Sixth column shows the relative height difference between TomoSAR point clouds and LiDAR data, as well as between TanDEM-X raw DEM and LiDAR data.}
\label{tab:statistics_comp9}
\begin{tabular}{c|c|cccc|cccc|c|c}
\toprule
\toprule
        \multirow{2}{*}{Structures} & \multirow{2}{*}{Sources} & \multicolumn{4}{c|}{Top} & \multicolumn{4}{c|}{Bottom} & \multirow{2}{*}{Height} & \multirow{2}{*}{Absolute Height Difference} \\
\cmidrule(lr){3-6} \cmidrule(lr){7-10}
      \multirow{2}{*}{} &   & Min  & Max & Std & Mean & Min  & Max  & Std  & Mean  &  \\
\midrule
\multirow{3}{*}{Structure 1} & T & -3.76 & 2.59 & 1.22 & -0.75 & -19.30 & -12.87 & 1.15 & -15.26 & 14.51 & \textbf{0.69}\\
                    & L &   -    &  -    &  -    & 539.01 & -     &  -    & -     & 525.19      &  13.82 & \textbf{-}\\
                    & D & 583.19 & 597.58 & 2.32 & 587.91 & 566.16 & 570.15 & 2.01 & 568.06 & 19.84  & \textbf{6.02}\\
\midrule
\multirow{3}{*}{Structure 2} & T & 20.39 & 22.62 & 0.56 & 21.39 & -27.85 & -22.62 & 1.18 & -25.30 & 46.70 & \textbf{0.75}\\
                    & L &   -    &  -    &  -    & 559.09 & -     &  -    & -     & 513.14      &  45.95 & \textbf{-}\\
                    & D & 598.57 & 642.43 & 8.35 & 624.99 & 556.61 & 583.46 & 4.45 & 574.83 & 50.16  & \textbf{4.21}\\
\midrule
\multirow{3}{*}{Structure 3} & T & 13.84 & 17.04 & 0.97 & 15.32 & -25.52 & -21.56 & 1.09 & -23.67 & 38.49 & \textbf{0.90}\\
                    & L &   -    &  -    &  -    & 552.97 & -     &  -    & -   & 515.38 & 37.59 & \textbf{-} \\
                    & D & 594.31 & 599.13 & 2.12 & 596.15 & 562.03 & 569.28 & 1.60 & 565.18 & 30.97  & \textbf{6.62}\\
\midrule
\multirow{3}{*}{Structure 4} & T & -4.61 & -1.90 & 0.74 & -2.91 & -13.84 & -10.35 & 0.83 & -12.05 & 9.14 & \textbf{0.67}\\
                    & L &   -    &  -    &  -    & 535.04 & -     &  -    & -   & 526.57 & 8.47 & \textbf{-}\\
                    & D & 582.41 & 584.94 & 0.84 & 584.06 & 572.76 & 574.96 & 0.48 & 573.42 & 10.64  & \textbf{2.17}\\
\midrule
\multirow{3}{*}{Structure 5} & T & -3.95 & -0.96 & 0.63 & -2.44 & -15.94 & -13.67 & 0.54 & -14.61 & 12.17 & \textbf{0.96}\\
                    & L &   -    &  -    &  -    & 535.62 & -     &  -    & -   & 524.41 & 11.21 & \textbf{-}\\
                    & D & 583.26 & 587.26 & 0.96 & 584.77 & 572.71 & 578.98 & 1.22 & 575.58 & 9.19  & \textbf{2.02}\\
\midrule
\multirow{3}{*}{Structure 6} & T & 10.42 & 13.47 & 0.49 & 12.11 & -16.05 & -14.31 & 0.82 & -15.61 & 27.72 & \textbf{0.67}\\
                    & L &   -    &  -    &  -    & 551.73 & -     &  -    & -   & 523.34 & 28.39 & \textbf{-}\\
                    & D & 587.23 & 594.29 & 2.12 & 589.76 & 567.22 & 570.82 & 1.32 & 569.36 & 20.4  & \textbf{7.99}\\
\midrule
\multirow{3}{*}{Structure 7} & T & 2.71 & 6.65 & 0.87 & 4.99 & -27.57 & -21.32 & 1.41 & -24.25 & 29.24 & \textbf{0.60}\\
                    & L &   -    &  -    &  -    & 548.01 & -     &  -    & -   & 519.37 & 28.64 & \textbf{-}\\
                    & D & 574.71 & 597.30 & 5.21 & 588.27 & 563.98 & 571.78 & 2.26 & 568.09 & 20.18  & \textbf{8.46}\\
\midrule
\multirow{3}{*}{Structure 8} & T & 0.06 & 6.42 & 1.34 & 4.06 & -20.96 & -20.27 & 0.18 & -20.69 & 24.75 & \textbf{0.94} \\
                    & L &   -    &  -    &  -    & 542.96 & -     &  -    & -   & 517.27 & 25.69 & \textbf{-}\\
                    & D & 584.73 & 598.09 & 3.36 & 590.40 & 566.26 & 574.70 & 2.62 & 570.12 & 20.28 & \textbf{5.41} \\
\midrule
\multirow{3}{*}{Structure 9} & T & -7.53 & -6.73 & 0.16 & -7.41 & -23.13 & -22.57 & 0.29 & -22.85 & 15.44 & \textbf{0.89}\\
                    & L &   -    &  -    &  -    & 530.39 & -     &  -    & -   & 515.84 & 14.55 & \textbf{-}\\
                    & D & 580.67 & 581.22 & 0.11 & 580.97 & 567.14 & 573.42 & 1.56 & 569.3 & 11.67  & \textbf{2.88}\\
\bottomrule
\bottomrule
\end{tabular}
\end{table*}

\subsection{Coregistration of different point clouds}
Consequently, when the TomoSAR point cloud is transformed to a UTM coordinate, its position may differ from the ground truth since the height of the reference point is unknown. Hence, the alignment of different point clouds is necessary. The most popular 3-D point cloud registration algorithm is iterative closest points (ICP) approach \cite{bib:zhang1994iterative}.

The performacne of ICP depends on the initial alignment. Hence, a coarse alignment is adopted before applying ICP, which includes three steps: (1) the edge image is extracted by an edge detector, such as Sobel algorithm \cite{bib:sobel2017isotropic}; (2) the horizontal coregistration of two edge images is using cross-correlation of two edge images; (3) the vertical coregistration is using cross-correlation of the two height histograms. After the coarse alignment, ICP can be applied for the fine alignment \cite{bib:wang2017fusing}.

\subsection{Object-based raster data generation}
The direct comparison of TomoSAR and LiDAR points is not feasible, as the central position of two corresponding points (one TomoSAR and one LiDAR point) and the footprint of the points are differing. Consequently, for comparing both data, an object-based raster need to be generated by using geographic information system GIS data.

\subsection{Comparison of individual structure}
In order to evaluate the estimation accuracy, nine test sites with high average SNR have been chosen for individual quantitative comparison. Fig. \ref{fig:optical_comp9} shows the optical images of nine test sites for quantitative comparison of NL-TomoSAR point clouds and TanDEM-X DEM. They are (1) Munich central station, (2) European bureau of patent, (3) Technical University of Munich, (4) A railway signal light stand near Hirschgarten, (5) A train repair garage near Hirschgarten, (6) A residential building between two bridges (Hackerbr{\"u}cke and Donnersbergebr{\"u}cke), (7) Munich University of Applied Sciences, (8) A residential building near Lowenbrau beer company and (9) Karstadt (shopping mall). The summary of the results is shown in Tab. \ref{tab:statistics_comp9}.

From Tab. \ref{tab:statistics_comp9} we can see that the height differences between TomoSAR result and LiDAR data are within one meter and the height differences between TanDEM-X DEM product and LiDAR data vary from 2.5 m to 8.5 m. Similar performance is shown in standard deviation, for NL-TomoSAR is up to 1.4 m and for TanDEM-X DEM is up to 8.4 m.

\subsection{Average accuracy}
\begin{figure}[!ht]
  \centering
  \includegraphics[width=0.5\textwidth]{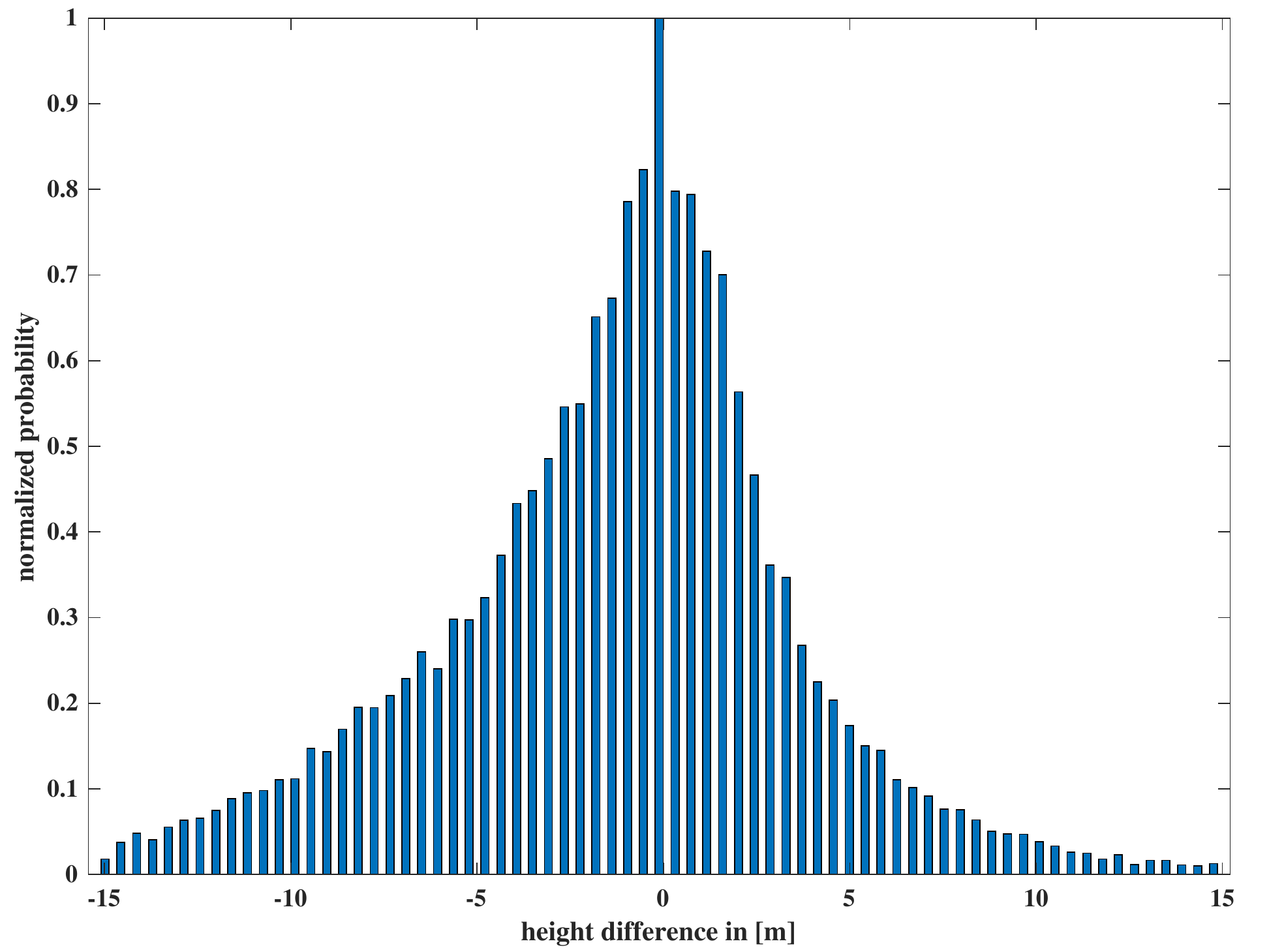}
  \caption{Histogram of height differences of structures in the whole Munich area.}
  \label{fig:comp_munich_hist}
\end{figure}

In order to have an assessment of the overall accuracy in a city scale, we compared all the 36,499 buildings in the area with the LiDAR point cloud. 38.7\% buildings are within 1 m accuracy. 62.8\% are within 2 m accuracy. A detailed distribution of accuracy is listed in Tab. \ref{tab:statistics_whole}. However, the two datasets (TanDEM-X CoSSC and LiDAR) were acquired at different time. It is almost certain that changes happened during the period. Therefore, in order to obtain a more realistic assessment, we truncated the distribution of height difference at $\pm \textrm{15 m}$. The truncated histogram can be seen in Fig. \ref{fig:comp_munich_hist}. 34,054 buildings remains after the truncation. Their overall standard deviation is 1.96 m.

\begin{table}[!ht]
\begin{center}
\caption{Statistics of quantitative comparison of the whole city}
\label{tab:statistics_whole}
\begin{tabular}{cc}
\toprule
Percentage of buildings &  Estimation accuracy \\
\midrule

38.7\% & within 1 m \\
62.8\% & within 2 m \\
93.3\% & within 15 m \\
\bottomrule
\end{tabular}
\end{center}
\end{table}

\section{Conclusion}
A new SAR tomographic inversion framework tailored for very limited number of measurements is proposed in this paper. A systematic investigation of the estimation accuracy of TomoSAR with a micro-stacks is carried out using simulated data. Our experiments show that SVD and CS-based methods have almost identical performance on the estimation of single scatterer and the SNR plays a more important role than $N$ for the estimation accuracy, when $N$ is small. For the estimation of double scatterers, CS-based approach outperforms the other spectral estimators. Experiments using TanDEM-X bistatic data shows the relative height accuracy of 2 m can be achieved in large scale. Thus it demonstrates the proposed framework being a promising solution for high quality large-scale 3-D urban mapping.

\end{document}